\documentclass[aps,prl,groupedaddress,showpacs]{revtex4}
\usepackage{graphicx}

\def\be{\begin{equation}}
\def\ee{\end{equation}}
\def\ba{\begin{eqnarray}}
\def\ea{\end{eqnarray}}
\def\bc{\begin{center}}
\def\ec{\end{center}}

\begin{document}

\title{Non-linear self-consistent response of graphene in time domain}

\author{S. A. Mikhailov}
\email[Electronic mail: ]{sergey.mikhailov@physik.uni-augsburg.de}

\affiliation{Institute for Physics, University of Augsburg, D-86135 Augsburg, Germany}

\date{\today}

\begin{abstract}
We study the non-linear electromagnetic response of graphene taking into account the self-consistent-field effects. Response of the system to a strong pulse excitation is calculated. It is shown that radiative decay in graphene differs from that of conventional two-dimensional electron systems both quantitatively and qualitatively. Possible applications of the predicted effects for generation of terahertz radiation are discussed.
\end{abstract}


\maketitle

Recently discovered \cite{Novoselov05,Zhang05} new two-dimensional (2D) material -- graphene -- attracted much attention in the past three years \cite{Katsnelson07,Geim07a}. Relativistic-like energy spectrum and the vanishing effective mass of charge carriers in graphene result in unconventional and interesting transport and electrodynamic properties, as has been demonstrated in numerous experimental and theoretical studies \cite{Novoselov07,Gusynin05,Ziegler06,Katsnelson06,Cheianov06,Nilsson06,Nomura07,Bostwick07,Deacon07,Gusynin06a,Falkovsky06,Sadowski06,Sadowski07,Gusynin06b,Gusynin07,Gusynin07a,Hwang06,Vafek06,Apalkov06,Abergel07,Ryzhii06,Wunsch06,Trauzettel07,Ryzhii07,Rana07,Mikhailov07d,Mikhailov07e}. 

As a direct consequence of the ``relativistic'' spectrum, electromagnetic response of graphene was predicted \cite{Mikhailov07e} to be strongly nonlinear: Irradiation of a graphene layer by electromagnetic wave with the frequency $\Omega$ should lead to the higher harmonics generation at the frequencies $m\Omega$, $m=3,5,7,\dots$, with the higher harmonics amplitudes  falling down very slowly, as $1/|m|$.

In general, considering response of charge carriers to the external electromagnetic radiation, one should take into account self-consistent-field effects \cite{Ehrenreich59}. The external time-dependent electric field of the wave ${\bf E}^{ext}(t)$ induces in the 2D gas, lying in the plane $z=0$, the electric current ${\bf j}(t)$. According to the Maxwell equations, this time-dependent current produces, in its turn, a secondary (induced) electric field 
\be
{\bf E}^{ind}_{z=0}(t)=-2\pi{\bf j}(t)/c, \label{EindJ}
\ee
which is added to the external field and acts back on the electrons. It is the total self-consistent electric field ${\bf E}_{z=0}^{tot}(t)={\bf E}^{ext}_{z=0}(t)+{\bf E}^{ind}_{z=0}(t)$ that should be written in the equations of motion for the current ${\bf j}(t)$, instead of the external field ${\bf E}^{ext}_{z=0}(t)$. 

In conventional 2D systems with the parabolic electron dispersion, taking into account the self-consistent-field effects leads to the following equation for the current (${\bf j}=-en_s{\bf v}$, $d{\bf v}/dt=-e{\bf E}_{z=0}^{tot}/m^\star$)
\be
\frac{d{\bf j}(t)}{dt}+\Gamma_{par} 
{\bf j}(t)=\frac{n_s e^2}{m^\star}{\bf E}^{ext}_{z=0}(t),\label{conven}
\ee
where $\Gamma_{par} \equiv 2\pi n_s e^2/m^\star c$ is the radiative decay rate \cite{Mikhailov04a}, $n_s$ is the 2D electron density, $m^\star$ is the electron effective mass, ${\bf v}$ is the hydrodynamic velocity of electrons, and the subscript $par$ reminds that Eq. (\ref{conven}) refers to the conventional 2D electrons with the parabolic energy dispersion. In (\ref{conven}) we have ignored the scattering due to impurities and phonons [the corresponding term $\gamma{\bf j}(t)$ can be added to the left-hand-side of (\ref{conven})]. In high-electron mobility samples the radiative decay rate $\Gamma_{par}$ substantially exceeds the impurities/phonons scattering rate $\gamma$, $\Gamma_{par}\gg\gamma$. In finite-size 2D electron-gas samples the plasmon-, the cyclotron (in a magnetic field), and/or the magnetoplasmon resonances can be observed; the linewidth of these resonances in high-electron-mobility samples is mainly determined by $\Gamma_{par}$ \cite{Mikhailov96a,Mikhailov04a}.

The electromagnetic response of graphene is strongly nonlinear \cite{Mikhailov07e}. In Ref. \cite{Mikhailov07e}, however, it was calculated ignoring the self-consistent-field effects. The goal of this Letter is to derive the corresponding equations of motion for the current ${\bf j}$ and to calculate response of the system to a pulse excitation ${\bf E}^{ext}(t)={\bf E}_{0}\tau\delta(t)$, where ${\bf E}_{0}$ and $\tau$ are the amplitude and the duration of the pulse. The self-consistent non-linear response of graphene to a harmonic external electric field will be discussed elsewhere \cite{SMAA}. 

Like in \cite{Mikhailov07e}, we consider a 2D electron gas with the ``relativistic'' massless energy spectrum $\epsilon_{{\bf p}\pm}=\pm V\sqrt{p_x^2+p_y^2}$ under the action of an external time-dependent electric field ${\bf E}^{ext}(t)$. Assuming  that the Fermi energy $\epsilon_F>0$ lies in the electron band and the temperature is small, $T\ll \epsilon_F$, we describe the graphene response using the kinetic Boltzmann equation in the collisionless approximation,
\begin{equation}
\frac{\partial f_{\bf p}(t) }{\partial t} -\nabla_{\bf p} f_{\bf p}(t) e {\bf E}^{tot}_{z=0}(t)=0.\label{BE}
\end{equation}
Here the sign $+$ ($-$) corresponds to the electron (hole) band, $V\approx 10^8$ cm/s is the effective ``velocity of light'' in graphene, and $f_{{\bf p}+}(t)\equiv f_{\bf p}(t)$ is the momentum distribution function of electrons (from now on we omit the sign $+$ for brevity). 
Equation (\ref{BE}) has the exact solution
\begin{equation}
f_{\bf p}(t) = {\cal F}_0\left({\bf p}-{\bf p}_0(t)\right),
\end{equation}
where ${\cal F}_0({\bf p})$ is the Fermi-Dirac function, and ${\bf p}_0(t)$ resolves the classical single particle equation of motion 
\be
\frac{d{\bf p}_0(t)}{dt}=-e{\bf E}^{tot}_{z=0}(t).\label{p0eq}
\ee
The electric current ${\bf j}(t)=-eg_sg_vS^{-1}\sum_{{\bf p}} {\bf v} f_{\bf p}(t)$ is written in terms of ${\bf p}_0(t)$ as
\begin{equation}
{\bf j}(t)= -\frac{g_sg_veV}{(2\pi \hbar)^2}\int dp_xdp_y \frac{{\bf p}}{p} {\cal F}_0\left({\bf p}-{\bf p}_0(t)\right),
\label{jx1}
\end{equation}
where $g_s=g_v=2$ are the spin and valley degeneracies in graphene, $p=\sqrt{p_x^2+p_y^2}$, and $S$ is the sample area. Combining Eqs. (\ref{p0eq}), (\ref{jx1}) and (\ref{EindJ}) we get the following equation of motion for the momentum ${\bf p}_0(t)$:
\be
\frac{d{\bf p}_0(t)}{dt}+\frac{e^2g_sg_vV}{2\pi \hbar^2c}\int dp_xdp_y\frac{{\bf p}}{p} {\cal F}_0\left({\bf p}-{\bf p}_0(t)\right)=-e{\bf E}^{ext}_{z=0}(t).\label{eqp0}
\ee
Equations (\ref{eqp0}) and (\ref{jx1}) describe the non-linear self-consistent response of graphene to an arbitrary external electric field ${\bf E}^{ext}_{z=0}(t)$ in the considered approximations. After the non-linear equation (\ref{eqp0}) is resolved with respect to the momentum ${\bf p}_0(t)$, the current ${\bf j}(t)$ can be found from (\ref{jx1}).

Now assume that the temperature is zero and that the external electric field ${\bf E}^{ext}(t)$ is directed along the $x$-axis. Then ${\bf p}_0=(p_0,0)$ and the dimensionless $x$-component of the momentum $P(t)\equiv p_0(t)/p_F$ is determined by the equation

\be
\frac{dP(t)}{dt}+\Gamma G(P) 
=-\frac{e}{p_F}E^{ext}_x(z=0,t),\label{eqp0a}
\ee
where $p_F=\epsilon_F/V$ is the Fermi momentum, 
\be
G(P)=\sqrt{1+P^2}\frac 2\pi\int_0^{\pi/2}\cos\theta d\theta\left(\sqrt{1+Q\cos\theta}-\sqrt{1-Q\cos\theta}\right),
 \ \ Q=\frac{2P}{1+P^2}\le 1, \label{funG}
\ee
and 
\be
\Gamma=\frac{g_sg_v}4\frac{e^2}{\hbar c}\frac{2\epsilon_F}{\hbar}=V\frac{e^2}{\hbar c}\sqrt{g_sg_v\pi n_s}.\label{raddecgraph}
\ee
The function $G(P)$ determines the electric current: 
\be
\frac j{-en_sV}=G(P).
\ee

In the linear-response regime, when the external electric field is so small, that $|P|=|p_0(t)|/p_F\ll 1$, the function $G(P)$ is linear, $G(P)\approx P$ at $|P|\ll 1$, and Eq. (\ref{eqp0a}) gives
\be
\frac{dP(t)}{dt}+\Gamma P(t)=-\frac e{p_F}E^{ext}_x(z=0,t).\label{linear}
\ee
Equation (\ref{linear}) is similar to (\ref{conven}), and one sees that the frequency $\Gamma$ has the physical meaning of the radiative decay rate in graphene in the linear regime. In contrast to $\Gamma_{par}$, $\Gamma$ is proportional to the square-root of the charge carrier density. For experimentally relevant $n_s$ the value of $\Gamma$ lies in the subterahertz range, $\Gamma/2\pi$(THz)$\approx 0.41\sqrt{n_s (10^{12}/{\rm cm}^{2})}$. 

\begin{figure}
\includegraphics[width=8.5cm]{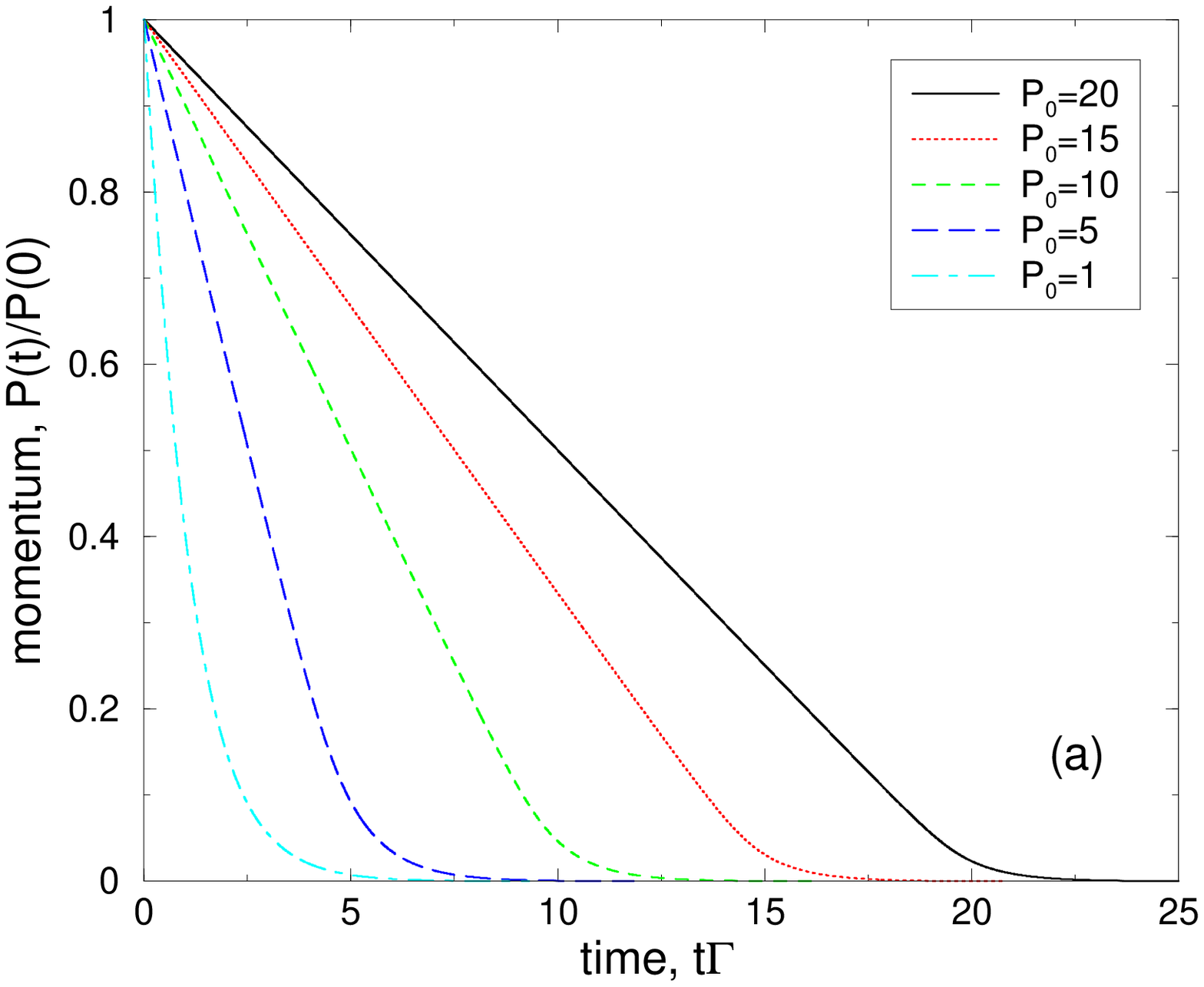}\\
\includegraphics[width=8.5cm]{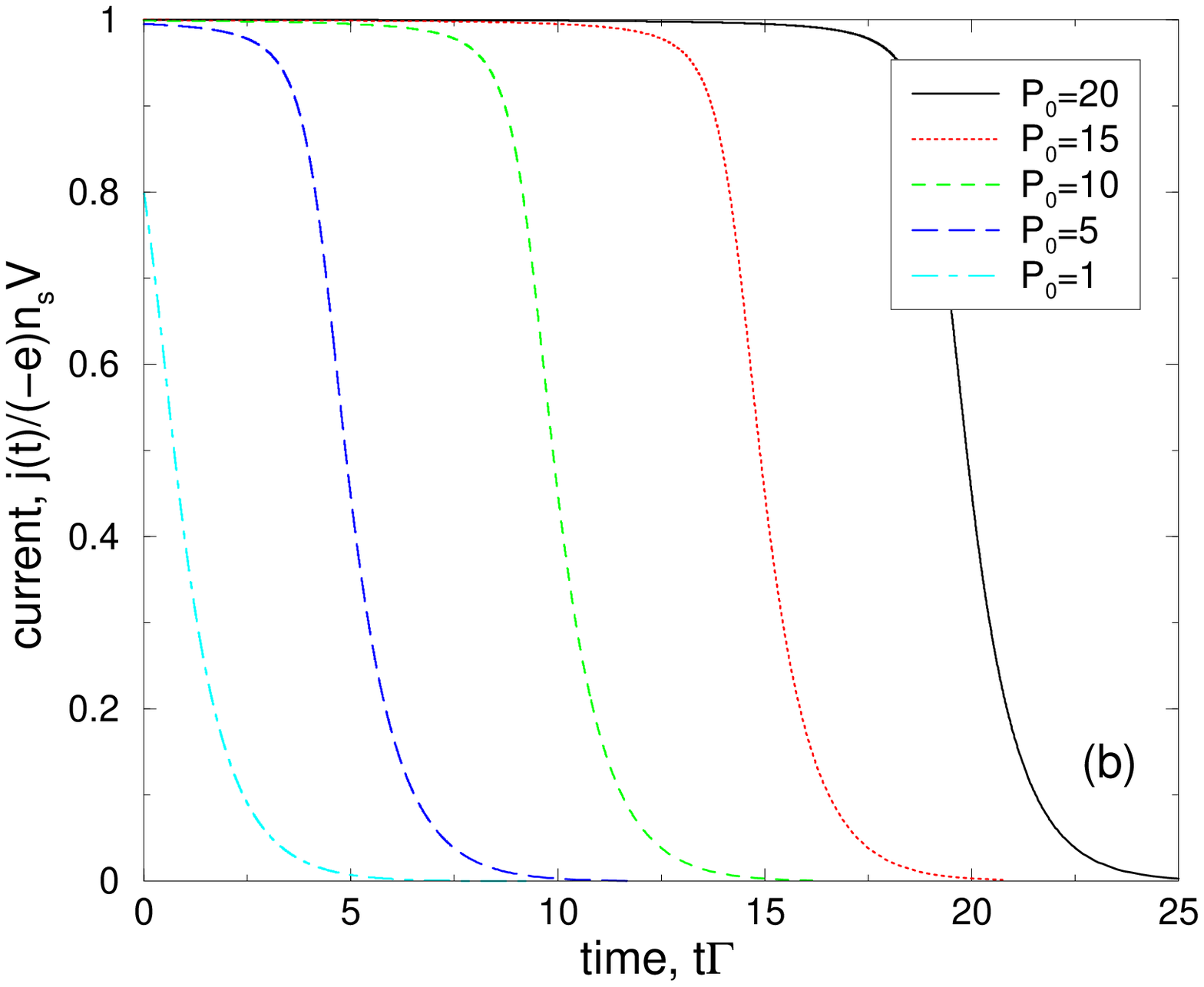}
\caption{\label{fig1} (Color online) The time dependence of {\bf (a)} the momentum $P(t)/P_0$ and {\bf (b)} the electric current $j(t)/(-e)n_sV$, at a pulse excitation of graphene. Different curves correspond to different pulse amplitudes $P_0=(-e)E_0\tau/p_F$.
}
\end{figure}

Now consider response of graphene to a pulse external electric field $E^{ext}_x(z=0,t)=E_0\tau\delta(t)$, not imposing any restriction on the field amplitude $E_0$. Eq. (\ref{eqp0a}) is reduced to the homogeneous differential equation 

\be
\frac{dP}{dt}+\Gamma G(P)= 0, \ \ t>0, \label{de}
\ee
with the boundary condition 
\be
P(t=+0)\equiv P_0=-\frac{eE_{0}\tau}{p_F}.\label{bc}
\ee
The problem (\ref{de})--(\ref{bc}) can be solved analytically.  
Expanding the integral in (\ref{funG}) in powers of $Q=2P/(1+P^2)\lesssim 1$, we get for the function $G(P)$:
\be
G(P)\approx \frac P{\sqrt{P^2+1}}\left[1+\frac 3{32}\left(\frac{2P}{P^2+1}\right)^2+\frac{35}{1024}\left(\frac{2P}{P^2+1}\right)^4\dots \right].
\label{G1}
\ee
The expansion (\ref{G1}) is valid both at $P\ll 1$ and $P\gg 1$; even at $P\approx 1$ Eq. (\ref{G1}) is very accurate: the third term is square brackets gives a correction smaller than 3.5\%. Integrating now Eq. (\ref{de}) with the boundary condition (\ref{bc}) we finally get the following implicit relation between the dimensionless momentum $P(t)$ and the time $t$:
\be
-\Gamma t \approx
\left(\sqrt{P'^2+1}-\ln\frac{1+\sqrt{P'^2+1}}{P'}+\frac {3/8}{(P'^2+1)^{1/2}}+\frac {13/96}{(P'^2+1)^{3/2}}-\frac{13/160} {(P'^2+1)^{5/2}}+\dots\right)_{P'=P_0}^{P'=P(t)}.\label{solP}
\ee
Figure \ref{fig1}a shows the dependence $P(t)$, given by Eq. (\ref{solP}), at different values of the electric field parameter $P_0=eE_0\tau/p_F$. If the external field is small, $P_0\ll 1$, the system relaxes after the pulse excitation exponentially, similar to the conventional 2D systems with the parabolic dispersion,
\be
P(t)=P_0\exp(-\Gamma t), \ \ P_0\lesssim 1. \label{expresp}
\ee
The characteristic decay time is determined in this case by the inverse radiation decay rate (\ref{raddecgraph}). If the external field is strong, $P_0\gg 1$, the momentum of the system decays linearly in time,
\be
P(t)=P_0-\Gamma t, \ \ P_0\gtrsim 1.\label{linresp}
\ee
The linear dependence (\ref{linresp}) remains valid until $P(t)$ reduces down to $P(t)\simeq 1$ (until $t\simeq P_0/\Gamma$); after that $P(t)$ decays exponentially like in (\ref{expresp}). 

The current $j(t)$ in the strong excitation regime $P_0\gg1$ is equal to its highest possible value $j_{max}=(-e)n_sV$ and time-independent at $t\lesssim P_0/\Gamma$, and then quickly (exponentially) decays (at the time scale $\simeq \Gamma^{-1}$) down to zero, Figure \ref{fig1}b.  

The fact, that after the pulse excitation electrons in graphene move with a constant velocity $V\approx 10^8$ cm/s for quite a long time $\sim P_0/\Gamma$, may have interesting applications. In a finite-size graphene sample such excited electrons will be reflected by the boundaries and oscillate in the sample with the typical frequency $\sim V/L$, lying in the terahertz range, if the sample dimensions $L\lesssim 1\ \mu$m. As at $P_0\gg 1$ the time $P_0/\Gamma$ is much longer than the oscillation period, this may lead to a coherent terahertz radiation from graphene excited by a strong pulse electric field.

To conclude, we have derived equations describing the non-linear self-consistent response of graphene electrons to an external time-dependent electric field and calculated response of graphene to a pulse excitation. We have shown that at low excitation strengths, the system responds exponentially, like conventional 2D electron layers, with however different characteristic decay rate. The radiative decay rate $\Gamma$ in graphene (\ref{raddecgraph}) is proportional to the square root of the electron density, in contrast to the normal 2D systems, where such a dependence is linear. At the strong excitation strengths, the average momentum of graphene electrons decays linearly, in contrast to the exponential decay in conventional systems, with the average current remaining constant during the time $\sim P_0/\Gamma$. The predicted effects may be used for terahertz generation.

This work has been partly supported by the Swedish Research Council and INTAS.

\bibliography{../../BIB-FILES/mikhailov,../../BIB-FILES/lowD,../../BIB-FILES/dots,../../BIB-FILES/graphene,../../BIB-FILES/emp}

\end{document}